\documentclass[11pt]{article}
\usepackage[latin1]{inputenc}
\usepackage[english]{babel}
\usepackage[namelimits]{amsmath}
\usepackage{amssymb}
\usepackage{amsmath}
\usepackage{amsthm}

\begin{document}
\title{A statistical representation of the cosmological constant from finite size effects at the apparent horizon}
\author{
Stefano Viaggiu,\\
Dipartimento di Matematica,
Universit\`a di Roma ``Tor Vergata'',\\
Via della Ricerca Scientifica, 1, I-00133 Roma, Italy.\\
E-mail: {\tt viaggiu@axp.mat.uniroma2.it}}
\date{\today}\maketitle
\begin{abstract}
In this paper we present a statistical description of the cosmological constant in terms of massless bosons
(gravitons). To this purpose, we use our recent results implying a non vanishing temperature 
${T_{\Lambda}}$ for the cosmological 
constant. In particular, we found that a non vanishing $T_{\Lambda}$ allows us to depict the cosmological constant 
$\Lambda$ as composed of elementary oscillations of massless bosons of energy $\hbar\omega$ by 
means of the Bose-Einstein distribution. In this context, as happens for photons in a medium,
the effective phase velocity $v_g$ of these massless excitations is not given by 
the speed of light $c$ but it is suppressed by a factor depending on the number of quanta present in the universe at the 
apparent horizon. We found interesting formulas relating the cosmological constant, the number of quanta $N$ and the mean value
$\overline{\lambda}$ of the wavelength of the gravitons. In this context, we study the possibility to look to the gravitons system 
so obtained as being very near to be a 
Bose-Einstein condensate. Finally, an attempt is done to write down the Friedmann flat equations in terms of 
$N$ and $\overline{\lambda}$.
\end{abstract}
{\bf Keywords} Cosmological constant . Gravitons . Apparent horizon . Bose-Einstein condensation .\\

\section{Introduction}
The standard concordant cosmological model is obtained by a spatially flat Friedmann metric endowed with a 
cosmological constant $\Lambda$ representing about $68\%$ of the present universe matter-energy content. 
Despite the enormous success of this model in explaining the main cosmological data, the real nature of this dark energy remains obscure and
a fundamental issue in modern cosmology. Many attempts have been done in the literature:
(see for example \cite{4}-\cite{r2} and references therein) quintessence, k-essence, phantom models, clock effects,
holographic dark energy, Bose-Einstein condensate (BEC), extended theories of gravity to cite someone of the most relevant.
In particular, the recent detection of gravitational waves \cite{r1} (GW150914 event) represents the born of the gravitational wave astronomy.
Gravitational wave astronomy, as noticed in \cite{r2}, opens the door to test general relativity against extended theories
of gravity (see \cite{r2} and references therein). In these extended theories a possible scalar component of the gravitational radiation
arises than can be verified by the study of the signal of gravitational waves. 
Another important issue concerning the cosmological constant $\Lambda$ is due to its very small value $\Lambda\simeq 10^{-52}/m^2$
(where '$m$' stands for meters), i.e. 
$10^{122}$ orders smaller than the value expected for vacuum energy in quantum field theory. Also a thermodynamic description of a
Friedmann universe endowed with a cosmological constant is a complicated and debated task \cite{22}-\cite{29}. 
First of all, the universe is a dynamical expanding system out from equilibrium. 
Moreover, it is not yet clear
how to describe \cite{29} the dynamical degrees of freedom related to the expansion of the universe.
As a consequence, a physically sound statistical mechanics description of the 
cosmological constant in terms of a fluid or gas is still lacking.
The knowledge of a thermodynamic description of the actual universe (at its 'thermodynamic radius', i.e. the apparent horizon
\cite{30,31} of our universe) is an important step for a better physical understanding of the cosmological constant.

Recently \cite{32,33,34,35}, we have generalized the Bekenstein-Hawking entropy formula suitable for black holes emebedded in 
Friedmann universes. In particular, our technology can be applied to the apparent horizon \cite{33,34,35} of Friedmann universes.
As a first important consequence \cite{33,34}, we have obtained $U_h=const.=0$. 
This can be interpreted with the fact that, according to an old conjecture \cite{36}, the
gravitational degrees of freedom encoded with an expanding universe are included in our tractation in such a way that the Misner-Sharp
energy $M_s c^2$ at the apparent horizon is exactly balanced by the negative gravitational expanding energy for a universe whose 
spatial sections are flat. Another important consequence of our setups is that \cite{35} the de Sitter universe, filled only with a 
non-vanishing positive cosmological constant $\Lambda$, is the only Friedmann solution that is in termal equilibrium with its 
sourronding. As a result, to a cosmological constant $\Lambda$ can be associated a non-zero temperature given by the one of the apparent horizon.
After introducing \cite{34} the Planck constant $\hbar$, a non vanishing internal energy $U_h$ with $U_h\sim T_{\Lambda}$
is allowed. All these facts permit us to explore a statistical description of $\Lambda$.

In section 2 we write down the first law at the apparent horizon for Friedmann spacetimes.
In section 3 we analyze the cosmological constant as composed of massless bosons (gravitons), 
while in section 4 the thermodynamic limit is discussed.
Finally, section 5 is devoted to some conclusions and final remarks.

\section{First law at the apparent horizon for Friedmann spacetimes}

A Friedmann spacetime is given in comoving coordinates by
\begin{equation}
ds^2=-c^2 dt^2+a^2(t)\left[\frac{{dr}^2}{1-kr^2}+r^2{d\Omega}^2\right],
\label{1}
\end{equation}
where $k=-1,0,+1$. The spacetimes (\ref{1}) filled
with a matter content satisfying the weak energy condition are equipped with an apparent horizon at the proper 
areal radius $L_h$ given by
\begin{equation} 
L_h=\frac{c}{\sqrt{H^2+\frac{k c^2}{a(t)^2}}}.
\label{2}
\end{equation} 
To the apparent horizon (\ref{2}) can be associated an holographic temperature (see for example \cite{14,15,16,20,21,22,25,26,29,30,31})
$T_h$, namely the Hawking temperature $T_h=c\hbar/(2\pi k_B L_h)$. In this paper we adopt the normalization used in
\cite{34,35} and the holographic temperature at the apparent horizon (\ref{2}) is 
\begin{equation} 
T_h=\frac{c\hbar}{4\pi k_B L_h}.
\label{3}
\end{equation} 
As well known \cite{26}, at the apparent horizon the Friedmann equations can be written in a form similar to the first law of thermodynamic
\begin{equation}
T_{h}dS=dU_h+W_h dV_h,\;\;\;W_h=\frac{p-\rho c^2}{2},
\label{4}
\end{equation}
where $W_h$ is the work term and $U_h=c^4L_h/(2G)$ is the Misner-Sharp energy term,
$\rho,p$ respectively the energy density and the pressure, 
provided that, as appens for static asymptotically flat black holes,
the apparent horizon is equipped with the entropy $S_h=\frac{k_B A_h}{4 L_P^2}$.
However, the Friedmann spacetimes are dynamical and non asymptotically flat and we expect, on general physical grounds, a modification for the
usual entropy law $S\sim A/4$. Moreover, as stated in \cite{27}, we have not a consistent way to calculate the 
dynamical degrees of freedom for a non-static gravitational field.

In \cite{32,33,34,35}, by using
suitable theorems for the formation of trapped surfaces in Friedmann
spacetimes, we proposed a new formula for the entropy of black holes embedded in Friedmann universes given by
\begin{equation}
S_{h}=\frac{k_B A_h}{4 L_P^2}+\frac{3k_B}{2c L_P^2}V_h H-
\frac{3k k_B}{4L_P^2}\frac{L_h V_h}{a(t)^2},
\label{5}
\end{equation}
where $V_h=4\pi L_h^3/3$ and $H$ the Hubble constant. In our proposal, thanks to the holographic principle, the 
(\ref{5}) is supposed to be the entropy of the whole universe at its apparent horizon where the entropy bound \cite{32,33,34,35}
is saturated as happens for the event horizon of a static black hole. The new first law at $L_h$ becomes \cite{33,34,35} the
(\ref{4}) but with
\begin{eqnarray}
& & dU_h=\frac{c^4}{2G}dL_h+\frac{c^3}{2G}L_h^2dH+\frac{kc^4}{2G}\frac{L_h^3}{a^3}da,\label{6}\\
& & W_h=-\frac{kc^4}{4\pi G a^2}+\frac{3 c^3 H}{8\pi G L_h}.\label{7}
\end{eqnarray}
A first consequence of (\ref{6}) is that, according to an old conjecture \cite{36}, the internal energy only for Friedmann flat solutions 
is vanishing \cite{33,34,35}\footnote{As shown in \cite{34}, when the Planck constant $\hbar$ is introducted, a non vanishing
	$U_h$ is allowed}.

Moreover \cite{35}, our new formulas (\ref{5})-(\ref{7}) imply that the only Friedmann solution that is in thermal
equilibrium with its sourronding is the de Sitter one, i.e. a Friedmann flat solution with a non-vanishing cosmological constant.
This means that the temperature of the apparent horizon (\ref{3}) is nothing else but the temperature of a de Sitter cosmological universe
$T_{\Lambda}$, i.e. $T_h=T_{\Lambda}$ for a de Sitter universe\footnote{This permit us to trace
back \cite{35} the thermal history of the universe in terms of $T_u-T_h$, being $T_u$ the temperature of the matter energy inside
$L_h$ where a de Sitter phase emerges when $T_u=T_h$}. This is in contradiction with the usual setups
(see \cite{27} and references therein) where a zero temperature is usually attributed to $\Lambda$. A non vanishing $T_{\Lambda}$
can be interpreted as a finite size effect. In fact, the Friedmann universe is thermodynamically a spherical object of areal radius
$L_h$. The usual results can be seen as the limit for $L_h\rightarrow\infty$ where $T_h\rightarrow 0$. Obviously, this finite volume effect
is negligible for our actual universe, but also a small but non-vanishing $T_{\Lambda}$ is conceptually important. In particular, as stated in
\cite{35}, the cosmological constant can be depicted as a fluid satisfying the equality
\begin{equation}
T_{h}^3 V_h = \frac{1}{48{\pi}^2}{\left(\frac{c\hbar}{k_B}\right)}^3,
\label{8}
\end{equation}
that looks like the equation of a reversible adiabatic transformation. Hence, according to this hypothesis \cite{35}, a fluid such that 
$T^3 V$ is less than the second member of (\ref{8}) is a phantom fluid.

The first law for the Friedmann flat case can be written as \cite{34}
\begin{equation}
T_{h}dS_h=dU_h+c^2\rho\;dV_h.
\label{10}
\end{equation} 
In the following two sections our reasonings do apply to a de Sitter universe equipped only with a positive non-vanishing
cosmological constant $\Lambda$ with $a(t)=e^{ct\sqrt{\frac{\Lambda}{3}}}, k=0$
and without dark matter and electromagnetic radiation. At the conclusions, section 5, we discuss the possible modifications of our results
in presence of dark matter and electromagnetic radiation.

Since a de Sitter expanding universe is in thermal equilibrium with its sourronding, we must have, for a cosmological constant with 
density ${\rho}_{\Lambda}$, $c^2{\rho}_{\Lambda}=|p_{\Lambda}|$: the apparent horizon is stationary ($L_h=const$) and the work term
$W_h$ must reduce to the usual expression $W=p\;dV$. 
The usual relation $p_{\Lambda}=-c^2{\rho}_{\Lambda}$, as noticed in \cite{29}, can be regained by the fact that the first law at
the apparent horizon has a different heat-sign convention with respect to the usual prescription.\\
We are now in the position to explore the possible statistical consequences of the relation (\ref{8}).

\section{Bose-Einstein distribution for the dark energy}

In this section we describe the cosmological constant of a de Sitter universe in terms of the usual Bose-Einstein distribution. 
To start with, we must evaluate the internal energy
of a de Sitter universe. As stated in \cite{33,35}, for all Friedmann flat spacetimes the internal energy $U_h$ is a constant of motion that for dimensional arguments (with only the constants $\Lambda,G,c$ at our disposal)
can be set to zero. In ordinary thermodynamics, to a vanishing internal energy can be associated a vanishing temperature. 
However, as stated above,
thanks to finite size effects at $L_h$ and introducing the Planck constant $\hbar$, a non vanishing temperature $T_{\Lambda}$ \cite{33,34}
satisfying the (\ref{8}) can be attributed to $\Lambda$. In \cite{34} we have also discussed the possible ultraviolet modification to the 
Friedmann flat equations caused by the introduction of the Planck constant.

In this paper we consider a de Sitter universe filled with the cosmological
constant $\Lambda$ composed of massless gravitons with a quanta of energy $\epsilon$ given by the usual relation
$\epsilon =\hbar\omega$. Concerning the chemical potential ${\mu}_{\Lambda}$, we may suppose that gravitons are interacting. We have a certain number
of interacting gravitons in thermal equilibrium within $L_h$. The situation is similar to the one of photons emitting a black body radiation in termal equilibrium
with a sourronding body. Hence, also in such a situation the number of gravitons is determined by the thermal equilibrium
condition for the free energy $F$ given by $dF=0$ and consequently ${\mu}_{\Lambda}=0$.\\ 
In a similar manner to the photons case, we can introduce
the wave number $\mathbf{k}$ and the number of oscillations in $d^3k=4\pi k^2 dk$ are given by
$V_h/{(2\pi)}^3 4\pi k^2 dk$. After introducing the usual relation $\omega=c k$ and since for gravitons we have only two independent 
polarizations corresponding to physical gravitons, as usual we have for the number of gravitons $N_h$ within $L_h=c/H_{\Lambda}$
\begin{equation}
N_h=\frac{V_h}{\pi^2 c^3}\frac{k_B^3T_h^3}{{\hbar}^3}\Gamma(3)\xi(3),
\label{11}
\end{equation}
where $\Gamma(3)=2$ and $\xi(3)\simeq 1.21$. It is easy to see that, thanks to (\ref{8}), 
$N=\Gamma(3)\xi(3)/(48{\pi}^4)<<1$. This result is reasonable only if we suppose a BEC of gravitons filling the ground state with 
$\epsilon=0$ at $T<T_h$. 
Considering gravitons as massless bosons, condensation cannot happen in ordinary situations \cite{37}.\\
Nevertheless, note that, thanks to the interactions among photons and the electrons of a medium, 
photons propagating in a medium with an effective phase velocity different from $c$. In fact, 
in a medium with index of refraction $\eta$, the effective phase velocity of a photon is $c/\eta$. In a similar way, we suppose that 
the interactions among gravitons physically justifies a slowing down of the gravitons phase velocity.
As a consequence, we may suppose that the following relation holds:
$\omega=\gamma |\mathbf{k}| c$, where the effective phase velocity $v_{g}$ is $v_{g}=c\gamma$. The parameter $\gamma$ must be fixed by
the overall Bose-Einstein statistics satisfied by gravitons. With the introduction of $\gamma$, the formula (\ref{11}) changes to
\begin{equation}
N_h=\frac{V_h}{\pi^2 c^3}\frac{k_B^3T_h^3}{{{\gamma}^3\hbar}^3}\Gamma(3)\xi(3).
\label{12}
\end{equation}
After using the equality (\ref{8}) characterizing the dark energy, we obtain for $\gamma$:
\begin{equation}
\gamma={\left(\frac{\Gamma(3)\xi(3)}{48{\pi}^4}\right)}^{\frac{1}{3}}\frac{1}{N_h^{\frac{1}{3}}}.
\label{13}
\end{equation}
Formula (\ref{13}) shows that the effective phase velocity of the gravitons depends only from the number of gravitons present at the horizon 
$L_h$ and is always smaller than $c$, a reasonable result. This result is a direct consequence of the constraint (\ref{8}).

To specify the system, we must evaluate the internal energy of the system at the apparent horizon. As stated in \cite{34},   
quantum fluctuations do imply a non-vanishing internal energy $U_h$. In this paper we consider only positive quantum fluctuations
\footnote{For the case of negative quantum fluctuations, a quantum field theory formalism is necessary \cite{38}.}.
In that case we can write \cite{34} $U_h=|c_0|c\hbar\sqrt{\Lambda}$, with $c_0$ a dimensionless constant. The total 
internal energy can be obtained, as 
usual, multiplying $dN_h$ for $\hbar\omega$ and integrating with respect to $\omega$: we obtain
\begin{equation}
U_h=\frac{V_h}{c^3}\frac{k_B^4T_h^4}{{{\gamma}^3\hbar}^3}\frac{{\pi}^2}{15}.
\label{14}
\end{equation}
After using $U_h=|c_0|c\hbar\sqrt{\Lambda}$ and equations (\ref{8}) and (\ref{14}) we have
\begin{equation}
U_h=k N_h c\hbar\sqrt{\Lambda},\;\;\;k=\frac{{\pi}^3}{60\sqrt{3}\Gamma(3)\xi(3)},
\label{15}
\end{equation}
i.e. $|c_0|\sim N$. Another interesting quantity is the mean value $\overline{\omega}$ of the frequency $\omega$. We get
\begin{equation}
\overline{\omega} = \frac{{\pi}^4 k_B T_h}{15\Gamma(3)\xi(3)\hbar}.
\label{16}
\end{equation} 
Hence, the following formula holds for $U_h$:
\begin{equation}
U_h=N_h \hbar\;\overline{\omega},
\label{17}
\end{equation}
which is the energy of $N$ oscillators with frequency $\overline{\omega}$.
Note that the mean frequency results independent on the number of gravitons $N_h$. However, this does not happens for the 
mean proper wavelength $\overline{\lambda}$ (denoting with ${\overline{\lambda}}_c$ the wavelength in comoving coordinates,
we have $\overline{\lambda}=e^{ct\sqrt{\frac{\Lambda}{3}}}\;{\overline{\lambda}}_c$). In fact we have 
$\overline{\lambda}=2\pi\gamma c/\overline{\omega}$ and as a consequence
\begin{equation}
\overline{\lambda}=\frac{120}{{\pi}^2}\Gamma(3)\xi(3)
{\left[\frac{\Gamma(3)\xi(3)}{48{\pi}^4 N_h}\right]}^{\frac{1}{3}}\sqrt{\frac{3}{\Lambda}}.
\label{18}
\end{equation}
The (\ref{18}) is an interesting formula relating the mean value for the wavelength of the gravitons, the cosmological constant and the number of excitations inside $L_h$. The (\ref{18}) can also be written in the form $\overline{\lambda}N_h^{1/3}/L_h\simeq 2.37$.  
Since $\Lambda\simeq 10^{-52}/{m}^2$, we have $\overline{\lambda}\simeq 2.37\times 10^{26}/(N_h^{1/3})\;m$.
This formula permits us some numerical investigations. As an example, for gravitons with $\overline{\lambda}\sim 10^{15} m$, we obtain
$N_h\sim 10^{33}$, while for $\overline{\lambda}\sim 10^6 m$ (binaries source) we obtain $N_h\sim 10^{60}$.

Concerning the mean frequency $\overline{\omega}$ given by (\ref{16}), for $\Lambda\sim 10^{-52}/{m}^2$ we obtain, thanks to the formula
$k_B T_h/\hbar=c/(4\pi)\sqrt{\Lambda/3}$, $\overline{\omega}\sim 10^{-18} Hz$, i.e. a very low frequency that we expect, for example, 
in a system near condensation.\\ 
In fact, thanks to the large number of gravitons $N_h>>1$ expected at the apparent horizon $L_h$ of a de Sitter universe\footnote{Practically of the same order of the one predicted by the concordance $\Lambda$CDM model at present 
	cosmological time.}, we have that $v_g<<c$ and for $N_h\rightarrow\infty$ we have
$v_g\rightarrow 0$. In the next sections we explore this possibility by presenting a suitable thermodynamic limit.

As a final consideration for this section, we study the thermalization of gravitons in an expanding universe.
Gravitons are expected to weakly interact among themselves. Moreover, the expansion of the universe can take more difficult thermalization.
Hence we expect that thermalizazion can certainly happen for a sufficiently slow expansion and for a huge gravitons number $N_h$.
The very low value for $\Lambda$ works in such a direction. To be more quantitative, we introduce the 
interaction rate $\Gamma$ representing the mean interactions frequency. As usual, we can assume that thermalization follows provided that
$\Gamma > H(t)=c\sqrt{\frac{\Lambda}{3}}$. For $\Gamma$ we could adopt the usual expression, i.e. 
$\Gamma=\frac{N_h}{V_h}{\sigma}_g {v}_g$, where ${\sigma}_g$ is the graviton-graviton cross section and 
${v}_g=c\gamma$. With the help of (\ref{12}) and (\ref{13}) we have:
\begin{equation}
N_h^{\frac{2}{3}} > {\left[\frac{\Gamma(3)\xi(3)}{48{\pi}^4}\right]}^{-\frac{1}{3}}\frac{4\pi}{\Lambda{\sigma}_g}. 
\label{c1}
\end{equation}
Since we have not a consolided quantum gravity theory on a curved spacetime, the cross section ${\sigma}_g$ in
(\ref{c1}) is left unspecified. Nevertheless, the (\ref{c1}) clearly shows that, also for a very small ${\sigma}_g$, a sufficiently 
large $N_h$ makes the job.

\section{Dark energy as an 'almost' Bose-Einstein condensate}

In the section above we have depicted the cosmological constant as composed of $N_h$ interacting gravitons with phase velocity
$v_g$ given by $v_g=c\gamma$, as happens for photons interacting with a given medium.
The adimensional factor $\gamma$ given by (\ref{13}) is calculated by the properties of the system at its thermodynamic radius, i.e. the apparent horizon $L_h=\sqrt{3/\Lambda}$. Since $\Lambda\sim 10^{-52}/{m}^2$ for our actual universe, we expect 
$N_h>>1$ and consequently $v_g\simeq 0$, i.e. the gravitons phase velocity slow down up to a value closed to zero, as expected for a
boson system
near the critical temperature.  
In particular, thanks to the (\ref{8}), we have for the temperature $T_{\Lambda}$ of
$\Lambda$, $T_{\Lambda}=T_h\sim 10^{-29}\;K$, i.e. a very cold temperature that is typical of BEC matter.
The mean frequency $\overline{\omega}$ is very small $\overline{\omega}\sim 10^{-18}\;Hz$ and the mean energy 
$\overline{\epsilon}$ of gravitons is of the order of $\overline{\epsilon}\sim 10^{-52}\;Joule$: this implies that the gravitons are
near the ground state with $\overline{\omega}\simeq 0$. In this regard, the temperature $T_h$, fixed thanks to the 
(\ref{8}), 
can be seen as a critical temperature.
Hence, the very low value for $\Lambda$ can well be a consequence of the fact that the cosmological constant is composed  of gravitons at 
very low temperatures, namely the 'critical' temperature $T_h$. The value of $T_h$ and the fact that
$\gamma <<1$ imply that effectively gravitons are very near to be in a BEC state.
In this regard, thanks to (\ref{8}) and (\ref{13}), 
gravitons at $T_{\Lambda}\simeq 0$ are practically frozen ($\overline{\omega}\simeq 0$ and $v_g\simeq 0$). 
As an example, for $N_h\sim 10^{60}$ (a relatively 'small' value), we have 
$v_g\sim c/10^{20}$, i.e gravitons employ $10^{15}$ seconds to travel one kilometer. We stress that
this almost BEC state of gravitons
at $T=T_h$ is due to the fact that $v_g\simeq 0$. Also note that, since for the critical 
temperature $T_c$ we have 
$T_c=T_{\Lambda}\simeq 0$, 
fluctuations allow to a non-zero fraction of gravitons to fill the state with $T<T_h$, therefore  with exactly
$\omega=0$. As a consequence of these reasonings, one may expect that gravitons at $T=T_h$ can acquire a kind of 
effective mass $m_g$. We can estimate this effective mass in the following way. From (\ref{16}) we see that the mean energy for graviton
$\overline{\epsilon}$ is given by
$\overline{\epsilon}=\hbar\overline{\omega}\simeq k_B T_h$. For a particle the relativistic dispersion relation is given by
$\overline{\epsilon}=\sqrt{c^2 p^2+m^2c^4}$. Since gravitons are frozen in a quasi BEC state we can assume $p=0$.
Hence we obtain  $m=\frac{\overline{\epsilon}}{c^2}\simeq 10^{-65} g$ for the effective mass acquired by gravitons.
The same result can be obtained by considering the effective mass as an effect due to the finiteness of the apparent horizon
$L_h\sim 10^{26}m$. We obtain $m_g=\hbar/(cL_h)\sim 10^{-65}g$, that is of the same order of the estimation above.
Note that also in \cite{37}, thanks to the 'finite' dye-filled optical microcavity where experiment is performed,
photons acquire an effective mass near the BEC ground state.

This phenomenon does not happen for huge values of the cosmological constant, for example at the primordial inflation 
$t_I\sim 10^{-37}-10^{-35}\;sec$,
where $\overline{\omega}>>0$ and $N_h$, thanks to the very small dimensions of the universe at the begin
of the primordial inflation, is expected to be certainly much more less than the actual expected value and as a consequence $v_g$ 
at $t=t_I$ is not negligible. Obviously, the tractation of section above is still valid for a finite value of $\Lambda$
expected at the primordial inflation.

In the following we perform the thermodynamic limit.
This a subtle issue since a Friedmann flat solution is spatially infinite. However, we have identified the apparent horizon 
$L_h$ with the thermodynamic radius of the system. Hence, according with this identification, the thermodynamic limit must be
performed by sending $N_h\rightarrow\infty$ and $L_h\rightarrow\infty$ (i.e. $\Lambda\rightarrow 0$) in such a way that
$\frac{N_h}{V_h}={\rho}_g=const$. In this way the cosmological constant $\Lambda$ 'disappears' in the thermodynamic limit,
but the gravitons density ${\rho}_g$ remains finite. This does not means that in the thermodynamic limit we have a minkowski
spacetime, but merely that a finite $\Lambda$ at a finite $L_h$ is spread on an infinite region. This is a mathematical procedure
to eliminate fluctuations depending on $N$ from the BEC phenomenology.

As a first consequence, in the thermodynamic limit above defined, we have $T_h=T_{\Lambda}\rightarrow 0$. From (\ref{16}) we deduce that
$\overline{\omega}\rightarrow 0$ an also the mean energy for gravitons $U_h/N_h\rightarrow$ with obviously
$U_h\rightarrow\infty$ (spatially infinite system). These features are typical of the BEC phenomenology. In particular, the 
mean frequency $\overline{\omega}$ must be vanishing in the ground state. Moreover, thanks to (\ref{13}) we have
$\gamma\rightarrow 0$ and as a result $v_g\rightarrow 0$. Finally, for $\overline{\lambda}$ we have
\begin{equation}
\overline{\lambda} \rightarrow \frac{1,47}{{{\rho}_g}^{\frac{1}{3}}}.
\label{19}
\end{equation}
The fact that the proper mean wavelength $\overline{\lambda}$ is not vanishing in the thermodynamic limit means that effectively the gravitons are not vanishing in this limit.\\ 
Summarizing, in the thermodynamic limit we have frozen ($v_g=0$) gravitons filling the ground state with zero mean energy and finite
mean wavelength: this implies that effectively exactly at $T_{\Lambda}=0$ gravitons are in a BEC state.

The actual very small value of the cosmological constant, thanks to our setups, is an indication that the identification of the actual
$\Lambda$ as composed of gravitons in a state very near to the ground state of a BEC condensate is a reasonable and viable possibility.

\section{Conclusions and final remarks}

In this paper, following our recent results \cite{32,33,34,35} concerning the thermodynamic of Friedmann universes,
we presented a statistical description of the cosmological constant. In particular, thanks to the fundamental result
(\ref{8}) that the cosmological constant has a non-vanishing temperature, we can describe $\Lambda$ as composed of
$N_h$ interacting gravitons with energy $\hbar\omega$ and $T_h=T_{\Lambda}$. The 
supposed interaction between gravitons justifies the 
assumption that their phase velocity slow down, as happens for photons propagating in a medium. The phase velocity $v_g$
is now $v_g=c\gamma$ with $\gamma\sim 1/N_h^{1/3}$.
As a consequence, we can calculate the mean frequency
for gravitons $\overline{\omega}$ and an interesting formula, namely the (\ref{18}), relating the mean wavelength
$\overline{\lambda}$, the cosmological constant $\Lambda$ and the gravitons number $N_h$ at the apparent horizon.
Thanks to the actual very low value for $\Lambda$, we obtain $\overline{\omega}\sim 10^{-18}$Hz. This very low value for
$\overline{\omega}$, together with the fact that practically $v_g\sim c/N_h^{1/3}\simeq 0$ motivates the view that the present day
dominant cosmological constant is constituted by gravitons very near a BEC state. In practice, the temperature
$T_{\Lambda}=T_h$ is a critical temperature for BEC and all the gravitons are very near the ground state with 
$\overline{\omega}=0$. The ground state $\overline{\omega}=0$ is rigorously obtained in the thermodynamic limit by sending
$L_h\rightarrow\infty$ ($\Lambda\rightarrow 0$) and $N_h\rightarrow\infty$ but with $N_h/V_h={\rho}_g$ held fixed. In this limit,
$\{\overline{\omega}, T_h, v_g\}\rightarrow 0$, with $\overline{\lambda}$ finite, motivating our physical interpretation
of the actual cosmological constant as composed of gravitons near the ground state of a BEC.

Gravitons are supposed massless. However, in massive gravity (see for example \cite{39} and references therein) a very 
small mass $m_g$ ($<10^{-62} g$) is allowed. Our tractation is still substantially valid by taking a small value for 
$m_g$ such that $m_g\leq \hbar\overline{\omega}/c^2$. For $\Lambda\sim 10^{-52}/{m}^2$, we have 
$m_g\leq 10^{-65}g$. 

Thes results of this paper have been obtained considering a de Sitter universe. Although our universe is composed by the $\simeq 68\%$
of dark energy, dark matter is also present ($\simeq 28\%$). It is thus interesting to wonder the possible modifications of the present paper 
calculations in presence of dark matter or other kinds of matter.\\ 
To this purpose, consider a Friedmann flat universe filled with a positive cosmological constant and
usual dust matter with density ${\rho}_m$ and radiation ${\rho}_r$ with temperature
$T_r$. First of all, it is rather physically reasonable to assume that gravitons decoupled from radiation during or before 
primordial inflation (see for example \cite{40} and references therein). With the actual $\Lambda$ made of gravitons, 
they are reasonable decoupled from the CMBR (and also dark matter starting from the recombination era) and as a consequence gravitons
cannot be thermalized with photons. Instead, we can reasonably assume that $T_g \sim T_{\Lambda}\sim T_h$ in our universe dominated by
$\Lambda$ and practically the results of sections 3-4 still hold in our Friedmann flat universe where matter, radiation and gravitons are
decoupled. In particular, gravitons composing $\Lambda$ are very near to a BEC state with 
$\{\overline{\omega}, v_g\}\simeq 0$.

As an useful example \cite{35}, we could consider ${\rho}_m$ and 
${\rho}_{\Lambda}=\Lambda c^2/(8\pi G)$ as a mixture
at the temperature $T_u>T_h=T_{\Lambda}$
\footnote{It is rather natural to suppose that $\Lambda$, thanks to (\ref{8}), is thermalized with the apparent horizon.} given by
\begin{equation}
T_u=\frac{{\rho}_{\Lambda}T_{\Lambda}+{\rho}_m T_m+{\rho}_r T_r}{{\rho}_{\Lambda}+{\rho}_m+{\rho}_r}.
\label{29}
\end{equation}  
In this case \cite{35}, $T_u>T_h$ and $L_{h,t}>0$. Only asymptotically $T_u\rightarrow T_h$ and a pure de Sitter phase emerges.
We can suppose again interacting 
gravitons with formulas (\ref{8}),(\ref{13}), (\ref{14}), (\ref{16}), (\ref{17}) still hold but with
$T_h=\frac{\hbar H(t)}{4\pi k_B}$, where $H(t)$ is the Hubble flow. Hence, the mean value $\overline{\omega}$ becomes time dependent
with $\overline{\omega}\sim H(t)$. When a de Sitter phase is reached with $H(t)\rightarrow c\sqrt{\Lambda/3}$ we regain the 
'stationary' formula (\ref{16}). Generally, we also expect that $N_h=N_h(t)$ and $\overline{\lambda}=\overline{\lambda}(t)$. 
By considering, for example,
a time independent internal energy $U_h$, we must have $N_h(t)\sim 1/H(t)$. Concerning formula (\ref{18}), we obtain
\begin{equation}
\overline{\lambda}(t) H(t) N_h^{\frac{1}{3}}(t)=2.37 c.
\label{30}
\end{equation} 
Note that, if we associate $N_h$ to the number of gravitons of $\Lambda$ at the apparent horizon, formula (\ref{30}) holds for
$N_h\neq 0$, i.e. for $\Lambda\neq 0$. 
However, with the hypothesis that gravitons are the 'quanta' of the gravitational field, we could 
suppose that formula includes gravitons constituing $\Lambda$ and the ones $N_{d_h}$ related to the intrinsic non static nature of a Friedmann 
universe, provided that they are thermalized with $\Lambda$, i.e. $T_{d_h}\simeq T_{\Lambda}$. In this case formula (\ref{30}) still 
holds with $N_h\rightarrow N_h+N_{d_h}=N$.\\
With respect to (\ref{30}), a de Sitter phase ($T_u=T_h=T_{\Lambda},\;L_{h,t}=0$) 
arises when $\overline{\lambda}(t)N^{\frac{1}{3}}(t)\simeq const$.
Stated in other words, by denoting with $\rho$ the density of the universe excluded $\Lambda$ and with $p$ the pressure with equation of state
$p=k\rho c^2$, Friedmann equations become  
\begin{eqnarray}
& &\frac{{\left(\overline{\lambda}(t)N^{\frac{1}{3}}(t)\right)}_{,t}}{\overline{\lambda}(t)N^{\frac{1}{3}}(t)}=
\frac{{\overline{\lambda}}_{,t}}{\overline{\lambda}}+\frac{N_{,t}(t)}{3N(t)}=
\frac{4\sqrt{3}\pi G\rho(1+k)}{\sqrt{8\pi G\rho+\Lambda c^2}},\label{31}\\
& &{\overline{\lambda}}^2N^{\frac{2}{3}}\left(\frac{8}{3}\pi G\rho+\frac{\Lambda c^2}{3}\right)=sc^2,\;\;s\simeq 5.61\label{32},
\end{eqnarray}
where only thermodynamic quantities are included, together with the (\ref{30}). 
For $\rho=0$ we have obviously the de Sitter solution with the relation (\ref{18}).
Note that 
for $k\rightarrow -1$ a de Sitter phase emerges: in this case 
we have ${\overline{\lambda}}^3(t)\sim 1/N(t)$ but with $\overline{\lambda}$ and $N$, differently from the de Sitter universe,
time dependent quantities.

As a final consideration, note that equation (\ref{18}) implies that the value of $\Lambda$ depends on the product 
${\overline{\lambda}}^2N_h^{\frac{2}{3}}$, i.e. $\Lambda\sim 1/({\overline{\lambda}}^2N_h^{\frac{2}{3}})$. 
Although with the formula (\ref{18}) we have not further constaints to obtain the actual very low estimated value for $\Lambda$, 
it is certainly true that a very low value is natural in our approach. For example, for wavelengths 
$\overline{\lambda} \geq 10^{6} m$, we have $N_h\leq 10^{60}$. In any case, thanks to the expected very large value for 
$N_h$, the issue of a very small $\Lambda$ can be certainly alleviated.
 
Summarizing, the thermodynamic (\ref{8}), (\ref{10}) at
the apparent horizon $L_h$ together with the hypothesis that dark energy is made of interacting
bosons very near to a BEC state due to massless gravitons or very
light bosons, is a physically viable possibility and must be certainly further investigated.

\end{document}